\documentclass[aps,pre,onecolumn,showpacs,amsmath,amssymb,longbibliography,notitlepage,11pt,tightenlines,superscriptaddress]{revtex4-1} %,twocolumn, pre or preprint
\usepackage[dvipsnames]{xcolor}
\usepackage{floatrow}
\usepackage{physics}
\usepackage{amsmath}
\usepackage{graphicx}
\definecolor{darkblue}{rgb}{0,0,0.6}
\definecolor{darkred}{rgb}{0.6,0,0}
\usepackage[colorlinks=true,urlcolor=darkblue,citecolor=darkblue, linkcolor=darkred,hyperfootnotes=false]{hyperref}

\newcommand{\Jij}{J_{ij}}

\begin{document}

\title{Materializing split, mixed, and three-body interactions using rotor-based mechanical hysterons}

\author{Oche~T.~Ali}
\affiliation{Department of Physics, St.~Olaf College, Northfield, MN 55057}

\author{Faten~Abu~Al~Ardat}
\affiliation{Department of Physics, St.~Olaf College, Northfield, MN 55057}

\author{Jack~Feider}
\affiliation{Department of Physics, St.~Olaf College, Northfield, MN 55057}

\author{Harry~Maakestad}
\affiliation{Department of Physics, St.~Olaf College, Northfield, MN 55057}

\author{Alex~Walk}
\affiliation{Department of Physics, St.~Olaf College, Northfield, MN 55057}

\author{Zachariah~S.~Schrecengost}
\affiliation{Department of Physics, Syracuse University, Syracuse, NY 13244}

\author{Joseph~D.~Paulsen}
\email{paulse8@stolaf.edu}
\affiliation{Department of Physics, St.~Olaf College, Northfield, MN 55057}

\begin{abstract}
Interacting hysteretic spins are an appealing model for cyclically-driven athermal disordered matter. 
Because they provide a basis for storing and processing information from their environment, such models are also being pursued as a framework for intelligent matter. 
Recent proof-of-concept designs have begun to demonstrate the strong, controlled, pairwise interactions that are necessary for this endeavor. 
But, it is not yet clear what are the limits---practically or fundamentally---on such interactions. 
Here we build rotor-based mechanical hysterons that extend the generality of their interactions in three ways: 
(i) splitting the interaction strength based on the hysteron state, 
(ii) building a non-reciprocal interaction of mixed sign, and 
(iii) incorporating tunable three-body effects. 
We access these effects within a simple, replicable design platform, and we rationalize our results. 
Our work expands the space of behaviors for designed structures that compute on mechanical inputs. 
\end{abstract}

\maketitle

\bigskip

\section{Introduction}
Disordered athermal matter is characterized by a multitude of locally stable states that grows at a breathtaking pace with the system size. 
Because such matter does not relax on its own to equilibrium, it naturally retains information about its past. 
Most striking is when memories of a global external drive are stored and can be read out by a simple means, such as the recollections of previous driving amplitudes in suspensions~\cite{Paulsen14,Chattopadhyay22,Padamata26}, amorphous solids~\cite{Fiocco14,Mungan19,Keim20,Shohat23a}, or crumpled films~\cite{Shohat22}. 
The study of memory formation thus offers a novel lens for understanding these complex materials~\cite{Keim19,Paulsen24,Kraus26}. 

Interacting hysteron models have emerged as a unifying model for a class of memory~\cite{Keim21,Lindeman21,Hecke21,Szulc22}. 
Each hysteron represents a two-state subsystem, which may be notional or real~\cite{Mungan19,Shohat22,Wang25}. 
Hysteron $i$ transitions between states $s_i = \pm 1$ when its local field passes switching threshold $\gamma_i^\pm$ with $\gamma_i^- < \gamma_i^+$ (Fig.~\ref{fig:1}a). 
Interactions between hysterons are tabulated in a matrix $\Jij$. 
These interactions may encourage ($\Jij>0$) or discourage ($\Jij<0$) one hysteron to match the state of another as follows: 
\begin{equation}\label{eq:model}
\gamma_i^\pm(S) = \gamma_{i,0}^\pm - \sum_{j} \Jij^\pm s_j \ ,
\end{equation}
where $\gamma_{i,0}^\pm$ are the bare switching fields absent interactions and $S=\{s_i\}$ is the system microstate. 
Tracking the behaviors of these hysterons under a time-varying drive vastly reduces the complexity of the original problem while providing testable hypotheses about the set of locally stable states and the pathways through them. 

Interacting hysteron models serve a second purpose. 
If Eq.~\ref{eq:model} can be materialized into artificial hysterons, it offers a design principle for building structures with memory and computation at their core~\cite{Paulsen24,Louvet25,Liu26}. 
Such systems process physical information sequentially~\cite{Liu24} and could be deployed in harsh environments or without a power source. 
And, because hysteresis is commonplace across different physics, this concept could be realized across different modalities including mechanical~\cite{Bense21,Jules22,Hyatt23,Liu24,El-Elmi24,Kamp25,Paulsen26}, pneumatic~\cite{Muhaxheri24,Djellouli24}, fluidic~\cite{Martinez-Calvo24,Rajput26} and electrical~\cite{Altman26}. 

Setting out on this path, one encounters several challenges. 
First, such embodied intelligence is only possible if a sufficiently broad set of switching thresholds and interactions can be accessed. 
Second, when building hysteron machines, additional deformation modes or behaviors can emerge that require additional hysteron or non-hysteron degrees of freedom~\cite{Ding22,Shohat25}. 
Third, even if the maps from physical parameters to switching thresholds and from switching thresholds to system behaviors are known, the corresponding inverse problems need to be addressed~\cite{Baconnier25,Teunisse25,Muhaxheri25,Teunisse26}. 
Thus, we do not yet understand the character or extent of the space of switching thresholds and interactions that can be realized in a physical system. 

Here we make progress by demonstrating a broadened set of interactions for a particular mechanical hysteron design. 
We work within the confines of mechanical rotors coupled with linear springs, which were introduced by Ref.~\cite{Paulsen26} and shown to have strong, controllable interactions of general sign that are faithful to Eq.~\ref{eq:model}. 
We first show how breaking symmetries of the rotor angles and spring mounting positions allows one to split $\Jij^+$ from $\Jij^-$. 
Then, after ruling out mixed interactions ($\Jij J_{ji}< 0$) over a range of rotor angles, we discover a principle for designing this exotic interaction, and we use it to arrange a pseudoavalanche. 
We then show how to introduce perturbative three-body interactions by extending the driving spring rest length. 
We close by considering symmetries in the rotor hysteron platform, providing a deeper explanation of these results. 
This work expands the space of physically-realizable abstract hysteron models in general, while providing concrete designs for this built system in particular.

\section{Hysteron design}

\begin{figure*}[p!]
\includegraphics[width=0.92\textwidth]{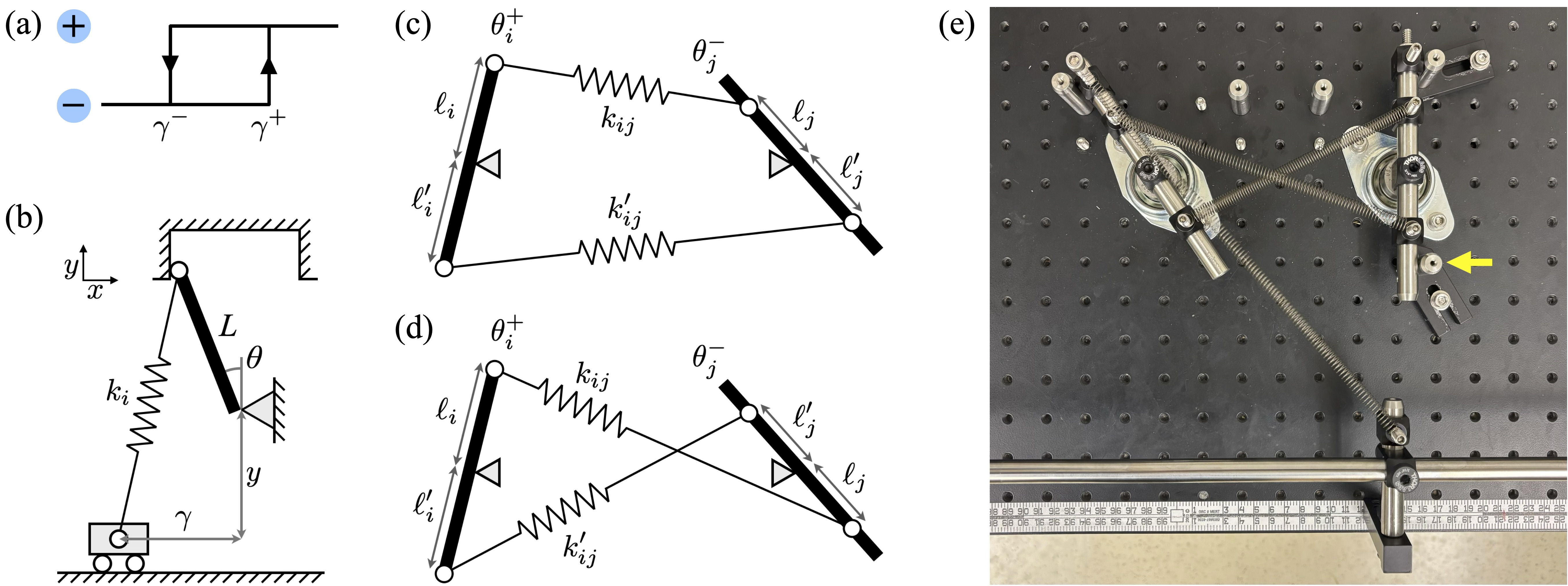}
\caption{
\textbf{Rotor hysteron design.}
(a) Abstract hysteron with two states and hysteretic transitions between them, $\gamma^- < \gamma^+$. 
(b) Following Ref.~\cite{Paulsen26}, we build a mechanical hysteron by coralling a hinged bar between rigid stoppers and penalizing intermediate rotor angles with a linear spring. 
We drive the system by translating a driving rod with displacement $\gamma$. 
The rotor is bistable when $\gamma^- < \gamma < \gamma^+$; outside this interval, there is one stable state. 
(c) We create tunable interactions by joining rotors $i$ and $j$ with coupling springs~\cite{Paulsen26}. 
Each spring exerts a torque $\tau_{ij}$ on rotor $i$. 
This uncrossed configuration creates a ferromagnetic-like interaction, $\Jij > 0$. 
(d) Crossed springs create an antiferrromagnetic-like interaction, $\Jij < 0$. 
(e) Experimental realization with two rotors. 
Here, $-\theta_1^- = -\theta_2^- = \theta_1^+ = 33.6^\circ$ and $\theta_2^+ = 0^\circ$. 
Arrow: An additional post holds $s_2=+$ while measuring switching thresholds for rotor $1$. 
}
\label{fig:1}
\end{figure*}

Following Ref.~\cite{Paulsen26}, we construct bistable rotors by mounting stainless steel bars on rotary ball bearings. 
Each bar is corralled between two posts that restrict its rotation to the angular interval $[\theta_i^- , \theta_i^+]$, where $\theta=0$ is along the $y$-axis and positive $\theta$ is in the clockwise direction so that $s_i=1$ corresponds to $\theta > 0$. 
Bistability is introduced by coupling the rotor to a long steel rod, using a linear spring of stiffness $k_i$ and rest length $x_0$, as schematized in Fig.~\ref{fig:1}b. 
This ``driving spring" attaches to the rotor a distance $L$ from its pivot, and the long steel rod (the ``driving rod") is located a distance $y$ from the rotor pivots. 
The driving rod is free to translate horizontally; its $x$-displacement serves as the global drive, $\gamma$. 
Finally, we mount driving spring $i$ to the driving rod with an offset $x_i$ such that this spring is parallel to the $y$-axis when $\gamma = -x_i$. 

Each rotor acts as a hysteron as we now describe. 
When the global drive $\gamma$ is less than a threshold value $\gamma_i^- = -y \tan \theta_i^+$, the rotor is pegged to the left post with $\theta_i = \theta_i^-$; we denote this as the $(-)$ state or $s_i = -1$. 
Likewise, when the global drive $\gamma > -y \tan \theta_i^- = \gamma_i^+$, the rotor is pegged to the right post in the $(+)$ state with $\theta_i = \theta_i^+$ and $s_i = 1$. 
Within the interval $\gamma_i^- < \gamma < \gamma_i^+$, the rotor is bistable and remains in the state it occupied previously. 

Reference~\cite{Paulsen26} showed how linking neighboring rotors together with additional springs leads to well-defined pairwise interactions, $\Jij$, that perturb the switching thresholds linearly following Eq.~\ref{eq:model}. 
Uncrossed springs produce ferromagnetic-like interactions (Fig.~\ref{fig:1}c) and crossed springs produce antiferromagnetic-like interactions (Fig.~\ref{fig:1}d). 

To bridge between the physical parameters of the rotors and springs and the quantitative interactions they produce, we need precise expressions for each torque in the setup, which were reported previously~\cite{Schrecengost25,Paulsen26}. 
We summarize these relations next.  
Throughout, we follow the convention that positive torques are in the clockwise direction, so that a spring mounted to rotor $i$ at location $\boldsymbol{x}_1$ with its other end at any $\boldsymbol{x}_2$ exerts a torque on rotor $i$ given by: 
\begin{equation}\label{eq:torque}
\tau = -\boldsymbol{r} \times \boldsymbol{F} = - k \boldsymbol{x}_1 \times \boldsymbol{x}_2 \left(1 - \frac{x_0}{ | \boldsymbol{x}_1 - \boldsymbol{x}_2 | } \right) \ ,
\end{equation}
where the origin is at the pivot of rotor $i$.

\section{Expressions for torques}

\subsection{Driving spring}
Rotor $i$ experiences a torque $\tau_i$ from a spring of stiffness $k_i$ that connects to the rotor a distance $L$ from its pivot and to the driving rod at $\boldsymbol{x}_2 = \{ \gamma + x_i , -y\}$. 
This torque may be calculated exactly using Eq.~\ref{eq:torque}:
\begin{equation}\label{eq:drive_exact}
\tau_i = L k_i \cos \theta_i ( \gamma + x_i + y \tan \theta_i ) \left( 1 - \frac{x_0}{\sqrt{ ( \gamma + x_i - L \sin \theta_i )^2 + (y + L \cos \theta_i)^2 }} \right) \ .
\end{equation}
When the rest length $x_0$ of the driving spring is small compared to $y+L$, Eq.~\ref{eq:drive_exact} simplifies to:
\begin{equation}\label{eq:drive_approx}
\tau_i \simeq L k_i \cos \theta_i ( \gamma + x_i + y \tan \theta_i ) \ , \ \ \text{when } x_0 \ll y + L \ . 
\end{equation}

\subsection{Single coupling spring}
We arrange multiple rotors along the $x$-axis spaced a distance $w$ apart. 
To mechanically couple nearby rotors we mount a spring of stiffness $k_{ij}$ and rest length $x_0$ to rotors $i$ and $j$ at signed distances $\ell_i$ and $\ell_j$ from their pivots. 
Equation~\ref{eq:torque} yields: 
\begin{equation}\label{eq:coupling_exact}
\tau_{ij} = k_{ij} \ell_i ( w \cos \theta_i + \ell_j \sin (\theta_j - \theta_i) ) \left( 1 - \frac{x_0}{\sqrt{ ( w + \ell_j \sin \theta_j - \ell_i \sin \theta_i )^2 + (\ell_j \cos \theta_j - \ell_i \cos \theta_i )^2 }} \right) \ .
\end{equation}
When the rest length $x_0$ is much shorter than the rotor spacing $w$, Eq.~\ref{eq:coupling_exact} simplifies to:
\begin{equation}\label{eq:coupling_approx}
\tau_{ij} \simeq k_{ij} \ell_i ( w \cos \theta_i + \ell_j \sin (\theta_j - \theta_i) ) \ , \ \ \text{when } x_0 \ll w \ . 
\end{equation}

\subsection{Pairs of coupling springs}
Except where otherwise stated, we use pairs of springs to couple the rotors and the symbol $\tau_{ij}$ represents the net torque on rotor $i$ from rotor $j$. 
One coupling spring is mounted at positions $\ell_i$ and $\ell_j$ to rotors $i$ and $j$, respectively, and the other is mounted at positions $\ell_i'$ and $\ell_j'$ (Fig.~\ref{fig:1}c,d). 
Uncrossed springs have $\ell_i \ell_j, \ell_i' \ell_j' > 0$, whereas crossed springs have $\ell_i \ell_j, \ell_i' \ell_j' < 0$.

\section{Thresholds and interactions}

Each rotor obeys torque balance,
\begin{equation}\label{eq:balance}
0 = \tau_i(\gamma,s_i) + \sum_{j \neq i} \tau_{ij} (s_i,s_j) + \tau_\text{post} \ ,
\end{equation}
where $\tau_\text{post}$ is the contact torque from one of the two posts corralling the rotor within the angular interval $[\theta^-,\theta^+]$. 
Varying $\gamma$ quasistatically, rotor $i$ switches its state when $\tau_\text{post} = 0$, as the driving spring makes this equilibrium unstable. 
Equation~\ref{eq:balance} therefore lets us calculate the two switching thresholds $\gamma_i^\pm(S)$ for each hysteron $i$ and each system microstate $S$. 

In this study, we either solve Eq.~\ref{eq:balance} numerically for $\gamma_i^\pm(S)$ using exact expressions for the torques, Eqs.~\ref{eq:drive_exact} and \ref{eq:coupling_exact}, or we will solve Eq.~\ref{eq:balance} analytically for $\gamma_i^\pm(S)$ if the approximate form for $\tau_i$ (Eq.~\ref{eq:drive_approx}) is used with either Eq.~\ref{eq:coupling_exact} or \ref{eq:coupling_approx} for $\tau_{ij}$. 
While sacrificing accuracy, the opportunity in this design platform to work with approximate analytical expressions is an advantage that we will use throughout this study. 

Of primary interest are the interactions, yet the thresholds are what we observe. 
Manipulation of Eq.~\ref{eq:model} shows that each interaction is a difference of thresholds:
\begin{equation}\label{eq:Jij_from_gamma}
\gamma_i^\pm(S |_{s_j=-}) - \gamma_i^\pm(S |_{s_j=+}) = 2 \Jij^\pm \ . 
\end{equation}
Equation~\ref{eq:Jij_from_gamma} thus gives direct access to the interactions, via pairs of thresholds that are measured in the experiment or calculated via torque balance (Eq.~\ref{eq:balance}). 

A special case arises when the rest length of the driving spring is sufficiently small. 
The approximate expression Eq.~\ref{eq:drive_approx} links $\gamma$ and $\tau_i$ via a piecewise affine relationship, such that the torque balance of Eq.~\ref{eq:balance} together with Eq.~\ref{eq:Jij_from_gamma} yields:
\begin{equation}\label{eq:Jij_from_torque}
\Jij^\pm \simeq \frac{ \tau_{ij}( s_i = \mp , s_j = +) - \tau_{ij}( s_i = \mp , s_j = -) }{2 L k_i \cos \theta_i^\mp} \ , \ \ \text{when } x_0 \ll y + L \ ,
\end{equation}
regardless of the states of all other hysterons.

\section{Splitting $\Jij^\pm$}\label{sec:Jplusminus}

The distinction between $\Jij^+$ and $\Jij^-$ is defined succinctly by Eq.~\ref{eq:Jij_from_gamma}, and Fig.~\ref{fig:2}a also illustrates the idea: $\Jij^+$ and $\Jij^-$ differ if the state of rotor $i$ affects its interaction with rotor $j$. 
Previous work in this system~\cite{Paulsen26} adopted several conditions on the parameters that led to the equality $\Jij^+ = \Jij^-$, so that a single matrix $\Jij$ describes the interactions. 
Namely, they focused systems where each pair of coupling springs is ``matched'' with equal spring constants and rest lengths ($k_{ij} = k'_{ij}$, $x_0 = x_0'$), and they are mounted at symmetrical mounting positions ($\ell_i = -\ell_i'$ and $\ell_j = -\ell_j'$).
In addition to some results at small rotor angles~\cite{Paulsen26}, they showed that at finite rotor angles, the equality $\Jij^+ = \Jij^-$ could be arranged exactly by using matched coupling springs and imposing mirror symmetry of the rotor angles, $\theta_i^- = -\theta_i^+$, even for arbitrary rotor angles and finite $x_0 = x_0'$. 

Although appealing in its simplicity, one wonders what additional control and tunability could be gained by abandoning these conditions. 
To answer this question, here we split the degeneracy of $\Jij^\pm$ systematically, first by breaking the symmetry of the rotor angles and then by breaking the symmetry of the spring mounting positions.

\subsection{Breaking the rotor angle symmetry}

We begin with experiments using uncrossed coupling springs.  
To isolate the effect of the rotor angles, we preserve the symmetry of the mounting positions: $\ell_1 = \ell_2 = -\ell_1' = -\ell_2'$. 
We keep the rest length of the driving spring small, $x_0/(y+L) \approx 0.1$. 
This choice is motivated by a desire to minimize any nonlinearities in the driving torque so that the results are more readily interpreted. 
Recalling that $\Jij^+$ measures how the $\gamma_i^+(s_j)$ thresholds are perturbed as $s_j$ switches states, we use the following protocol: (1) we clamp rotor $2$ in the $(+)$ state and drive rotor $1$ to measure $\gamma_1^\pm(+)$ as shown in Fig.~\ref{fig:1}e, (2) we clamp rotor $2$ in the $(-)$ state and drive rotor $1$ to measure $\gamma_1^\pm(-)$, (3) we take differences to calculate $J_{12}^\pm$ via Eq.~\ref{eq:Jij_from_gamma}. 

\begin{figure*}[p!]
\includegraphics[width=1.0\textwidth]{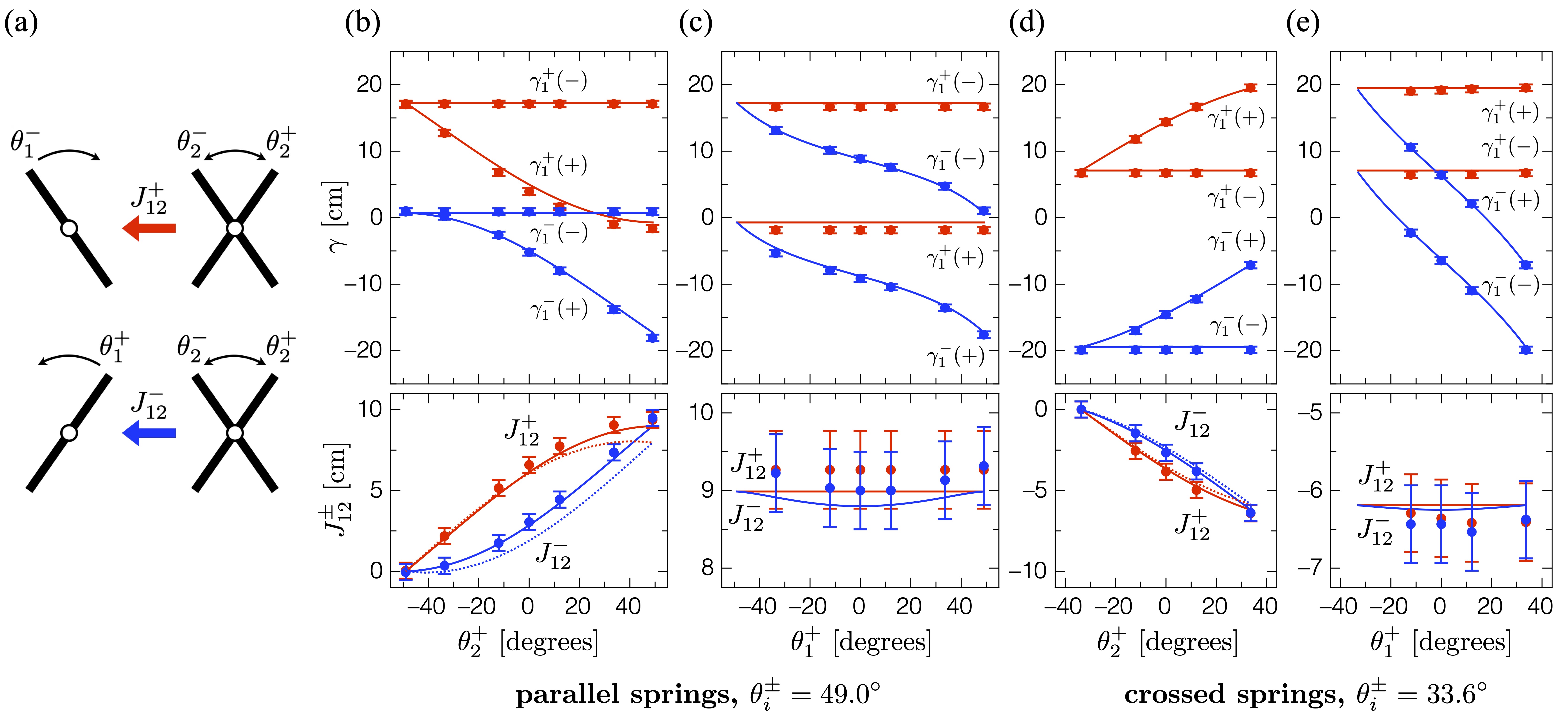}
\caption{
\textbf{Splitting $\Jij^\pm$ via rotor angles.}
(a) $J_{12}^+$ is the interaction from rotor $j$ when $\theta_i = \theta_i^-$. 
(b) Switching thresholds for rotor 1, which is connected to rotor 2 with uncrossed coupling springs. 
We break a symmetry in the system by varying $\theta_2^+$ away from $49.0^\circ$; this splits $J_{12}^\pm$. 
(c) Varying $\theta_1^+$ also breaks the symmetry, and splits $J_{12}^\pm$ to a lesser degree. 
(d,e) Analogous results for crossed springs with the maximum angles reduced to $\pm33.6^\circ$. 
Dashed lines in panels (b) and (d) are exact results for vanishing spring rest lengths, Eq.~\ref{eq:Jij_small-x0_matched}. 
There is no splitting of $\Jij^\pm$ in this limit for panels (c) and (e). 
Parameters for all panels: $y=15.0$ cm, $L=6.3$ cm, $w=15.24$ cm. 
Coupling springs: $k_{12}=0.19$ N/cm, $x_0=4.1$ cm, $\ell = \pm3.5$ cm. 
Driving spring: $k_1=0.070$ N/cm, $x_0 = 2.25$ cm. 
}
\label{fig:2}
\end{figure*}

In the first set of experiments, we vary $-49.0^\circ < \theta_2^+ < 49.0^\circ$ while holding fixed $-\theta_1^- = -\theta_2^- = \theta_1^+ = 49.0^\circ$. 
Figure~\ref{fig:2}b shows measurements of the four switching thresholds for rotor 1: $\gamma_1^\pm(+)$ and $\gamma_1^\pm(-)$, as $\theta_2^+$ is varied. 
We note that two of the switching thresholds are constant; this merely reflects the fact that $\theta_2^+$ is not involved in those transitions. 
Experimental uncertainties are assessed by measuring the same threshold multiple times and are found to be $\pm 0.5$ cm. 

We take differences to calculate $J_{12}^\pm$ via Eq.~\ref{eq:Jij_from_gamma}, also plotted in Fig.~\ref{fig:2}b. 
The results show an unambiguous splitting of $J_{12}^\pm$ as $\theta_2^+$ is decreased away from the symmetrical state, $\theta_2^+ = -\theta_2^-= 49.0^\circ$. 
As $\theta_2^+$ approaches $\theta_2^- = -49.0^\circ$, $J_{12}^+$ and $J_{12}^-$ must both go to zero for the simple reason that the two states for rotor 2 now coincide. 
We probe this behavior over finer increments of $\theta_2^+$ by solving for the exact switching thresholds numerically using Eqs.~\ref{eq:drive_exact}, \ref{eq:coupling_exact}, and \ref{eq:balance}. 
Figure~\ref{fig:2}b shows how the thresholds and interactions vary continuously. 
The numerical results support the experimental findings, which together demonstrate an appreciable, tunable splitting of $J_{12}^-$, as the symmetry $\theta_2^+ = -\theta_2^-$ is broken. 

We can also split $J_{12}^\pm$ by varying $\theta_1^+$. 
Figure~\ref{fig:2}c shows the four thresholds $\gamma_1^\pm(+)$ and $\gamma_1^\pm(-)$, as $\theta_1^+$ is varied and the resulting $J_{12}^\pm$. 
Here the splitting of $J_{12}^\pm$ is much weaker and exists at the level of the experimental uncertainties. 
The numerical results make us confident that there is a splitting, and we explain its weak magnitude later on. 

Next, we configure the coupling springs into a crossed configuration~\cite{Paulsen26}. 
Crossed springs increase the hysteresis of each rotor, so that $\gamma^+_1(s_2) - \gamma^-_1(s_2)$ becomes larger. 
In order to keep within the linear regime of the driving spring, we thus reduce the maximum rotor angles. 
We vary $-33.6^\circ < \theta_2^+ < 33.6^\circ$ while holding fixed $-\theta_1^- = -\theta_2^- = \theta_1^+ = 33.6 ^\circ$. 
Figure~\ref{fig:2}d shows that $J_{12}^\pm$ is split by varying $\theta_2^+$. 
The splitting is smaller than in Fig.~\ref{fig:2}b; this is expected due to the smaller rotor angles. 
Figure~\ref{fig:2}e shows that $J_{12}^\pm$ is are split slightly by varying $\theta_1^+$; once again we rely on the numerics to resolve this. 

\subsection{Limit of small spring rest lengths} 

To rationalize these results, we consider an analytically tractable limit in which all spring rest lengths are negligible. 
Analyzing Eqs.~\ref{eq:coupling_approx} and~\ref{eq:Jij_from_torque}, we find that for one coupling spring,
\begin{equation}\label{eq:Jij_small-x0_one}
\Jij^\pm \simeq  (2 L k_i \cos \theta_i^\mp )^{-1} k_{ij} \ell_i \ell_j \left[ \sin( \theta_j^+ - \theta_i^\mp ) - \sin( \theta_j^- - \theta_i^\mp ) \right] \ \ \ \text{(one spring)} \ . 
\end{equation}
We emphasize that Eq.~\ref{eq:Jij_small-x0_one} is valid for arbitrary rotor angles $\theta_i^\pm$, $\theta_j^\pm$. 
Intriguingly, including a second coupling spring merely replaces the factor $k_{ij} \ell_i \ell_j$ in Eq.~\ref{eq:Jij_small-x0_one} with $(k_{ij} \ell_i \ell_j + k'_{ij} \ell'_i \ell'_j)$, which does not access any new values for the interactions. 
Thus, when all the spring rest lengths are negligible, the interactions for one coupling spring and two coupling springs are readily mapped onto one another. 
We nevertheless use matched pairs of coupling springs whenever possible (i.e., $k_{ij} = k'_{ij}$, $\ell_i = -\ell_i'$, $\ell_j=-\ell_j'$), as this simplifies the behaviors when the rest lengths are finite~\cite{Paulsen26}. 
When the coupling springs are matched,
\begin{equation}\label{eq:Jij_small-x0_matched}
\Jij^\pm \simeq  (L k_i \cos \theta_i^\mp )^{-1} k_{ij} \ell_i \ell_j \left[ \sin( \theta_j^+ - \theta_i^\mp ) - \sin( \theta_j^- - \theta_i^\mp ) \right] \ \ \ \text{(matched springs)} \ ,
\end{equation}
with the sign of $\ell_i \ell_j$ encoding whether they are crossed. 

As we now show, Eq.~\ref{eq:Jij_small-x0_matched} captures the splitting of $\Jij^\pm$. 
We plot Eq.~\ref{eq:Jij_small-x0_matched} in Fig.~\ref{fig:2}b using the protocol for those data where we vary only $\theta_2^+$. 
The data match the trends of the experiments and the exact kinematic model, thereby identifying the angular dependence of Eq.~\ref{eq:Jij_small-x0_matched} as the dominant effect. 
Likewise, we plot Eq.~\ref{eq:Jij_small-x0_matched} in Fig.~\ref{fig:2}d where the springs are now crossed and we find good agreement. 

The muted dependence of $J_{12}^\pm$ on rotor $1$ in Figs.~\ref{fig:2}c,e can now be understood using Eq.~\ref{eq:Jij_small-x0_matched}. 
Setting $\theta_2^- = -\theta_2^+$ and using angle sum formulas, we find $J_{12}^+ = J_{12}^-$, regardless of the values of $\theta_1^\pm$. 
Thus, the weak splitting of $J_{12}^\pm$ and $J_{12}^-$ in Figs.~\ref{fig:2}c,e is a higher-order effect that relies on finite spring rest lengths. 

Equation~\ref{eq:Jij_small-x0_matched} also shows that asymmetric spring mounting positions should not split $\Jij^\pm$ to a large degree. 
Indeed, setting $\theta_i^- = -\theta_i^+$ and $\theta_j^- = -\theta_j^+$ in Eqs.~\ref{eq:Jij_small-x0_one} and~\ref{eq:Jij_small-x0_matched} yields $\Jij^+ = \Jij^-$, for arbitrary $\ell_i, \ell_j$. 
We now examine how finite spring rest lengths can change this picture. 

\subsection{Breaking the spring mounting position symmetry}

\begin{figure}[p!]
\includegraphics[width=0.45\textwidth]{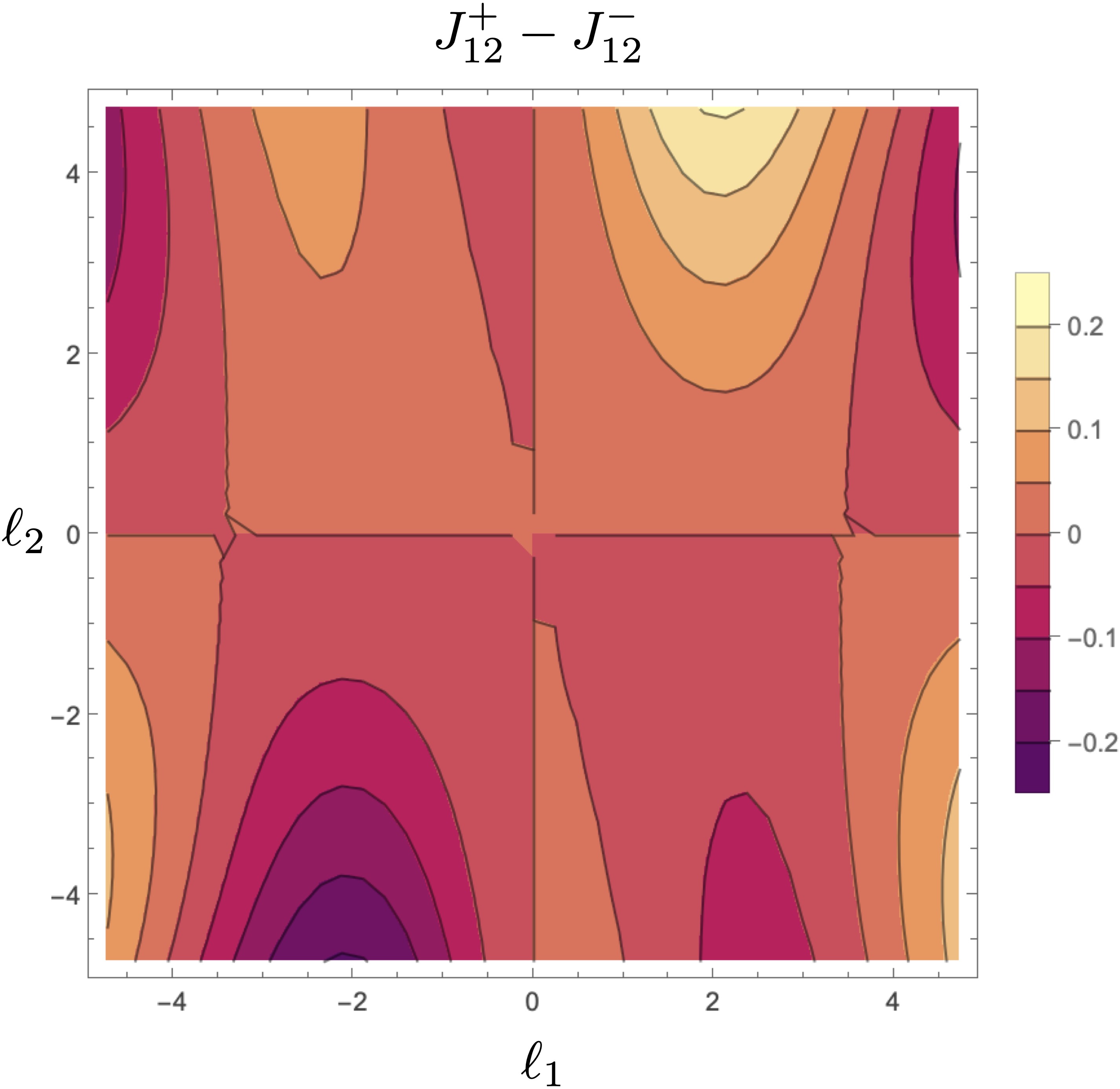}
\caption{
\textbf{Splitting $\Jij^\pm$ via spring mounting positions.}
Two rotors are coupled by a single spring mounted at positions $\ell_1$ and $\ell_2$. 
We plot $J_{12}^+ - J_{12}^-$ as a function of $\ell_1$ and $\ell_2$. 
Splitting of $\Jij^\pm$ is present but weak. 
In the limit of small spring rest lengths, $J_{12}^+ - J_{12}^-$ would vanish everywhere. 
Parameters: $y=15.0$ cm, $L=6.3$ cm, $w=15.24$ cm, $\theta_i^\pm = \pm33.6^\circ$. 
Coupling spring: $k_{12}=0.19$ N/cm, $x_0=4.1$ cm. 
Driving spring: $k_1=0.070$ N/cm, $x_0 = 2.25$ cm. 
}
\label{fig:3}
\end{figure}

We consider a single spring connecting two rotors at positions $\ell_1$ and $\ell_2$, with the symmetry $\theta_i^- = - \theta_i^+$ restored. 
We explore this parameter space numerically using the exact expressions for all torques. 
Figure~\ref{fig:3} plots the difference, $J_{12}^+ - J_{12}^-$, as both $\ell_1$ and $\ell_2$ are varied, while maintaining the remaining experimental parameters from the previous section. 
The results show that the coupling spring mounting positions do split $\Jij^\pm$ for finite $x_0$, although the effect is relatively weak. 
When we take the driving spring rest length to zero, the values of $J_{12}^+ - J_{12}^-$ fall to within machine precision of $0$ over the same domain. %$\mathcal{O}(10^{-15})$ 

\section{Mixed interactions}\label{sec:mixed}

So far we have considered ferromagnetic-like and antiferromagnetic-like interactions, but a third type of interaction can be theorized. 
In so-called ``mixed'' interactions, a hysteron exerts a ferromagnetic-like influence while receiving an antiferromagnetic-like influence from the other. 
This exotic interaction was considered in abstract hysteron models~\cite{Hecke21}, and Shohat \& van Hecke showed how to design it into networks of bistable springs~\cite{Shohat25}. 
However, we are not aware of any experimental demonstration of mixed interactions to date. 

It might seem unlikely that mixed interactions could be realized for the rotor hysterons described here. 
The simple pattern by which uncrossed springs lead to ferromagnetic-like interactions and crossed springs lead to antiferromagnetic-like interactions (via the sign of $\ell_i \ell_j$ in Eqs.~\ref{eq:Jij_small-x0_one} and~\ref{eq:Jij_small-x0_matched}) seems to rule it out. 
Motivated by this, Schrecengost proved the absence of mixed interactions for small spring rest lengths and symmetric rotor angles~\cite{Schrecengost25}. 
Here we extend this result to any asymmetric rotor angles within $\pm 45^\circ$, in the same small-$x_0$ limit. 
Then, to our surprise, we find that this bound is tight: Moving just one of the rotor angles beyond $45^\circ$ can produce a mixed interaction. 
We then pinpoint the geometric origin of this oddity. 
Equipped with controllable mixed interactions, we design a pseudoavalanche~\cite{Hecke21} and  realize it in the lab. 

\subsection{No mixed interactions for small $x_0$ and $|\theta| \leq 45^\circ$}

We first seek a clear parameter regime that demonstrably lacks mixed interactions. 
Working in the limit of small $x_0$, we evaluate the product $\Jij^\pm J_{ji}^\pm$ using Eq.~\ref{eq:Jij_small-x0_matched}. 
This product consists of a prefactor $(L k_i \cos \theta_i^\mp )^{-1} (L k_j \cos \theta_j^\mp )^{-1} k_{ij}^2 \ell_i^2 \ell_j^2$ that is manifestly non-negative, multiplied by an angle function:
\begin{equation}
f(\theta_i^-,\theta_i^+,\theta_j^-,\theta_j^+) \simeq \left[ \sin( \theta_j^+ - \theta_i^\mp ) - \sin( \theta_j^- - \theta_i^\mp ) \right] \left[ \sin( \theta_i^+ - \theta_j^\mp ) - \sin( \theta_i^- - \theta_j^\mp ) \right] \ .
\end{equation}
An analogous angle function $g$ is readily constructed for $\Jij^\pm J_{ji}^\mp$. 
The interactions are thus not mixed if $f, g > 0$. 
Both $f$ and $g$ consist of two factors of the form $[\sin(a-c) - \sin(b-c)]$ with $a-c > b-c$. 
Because $\sin(x)$ is monotone increasing for $-90^\circ \leq x \leq 90^\circ$, this structure guarantees that each factor is non-negative when $|\theta| \leq 45^\circ$. 
Mixed interactions are thus ruled out for small $x_0$ and $|\theta| \leq 45^\circ$.

\subsection{Mixed interactions using $|\theta| > 45^\circ$}

Given this result, it is natural to search for examples of mixed interactions at larger angles. 
One positive result is given by the following: $\theta_1^\pm = \pm 45^\circ$, $\theta_2^- = 45^\circ < \theta_2^+ < 90^\circ$, for which $f = -1 + \sin(45^\circ + \theta_2^+) < 0$ and $J_{12}^+ J_{21}^+ < 0$. 

\begin{figure}[p!]
\includegraphics[width=0.48\textwidth]{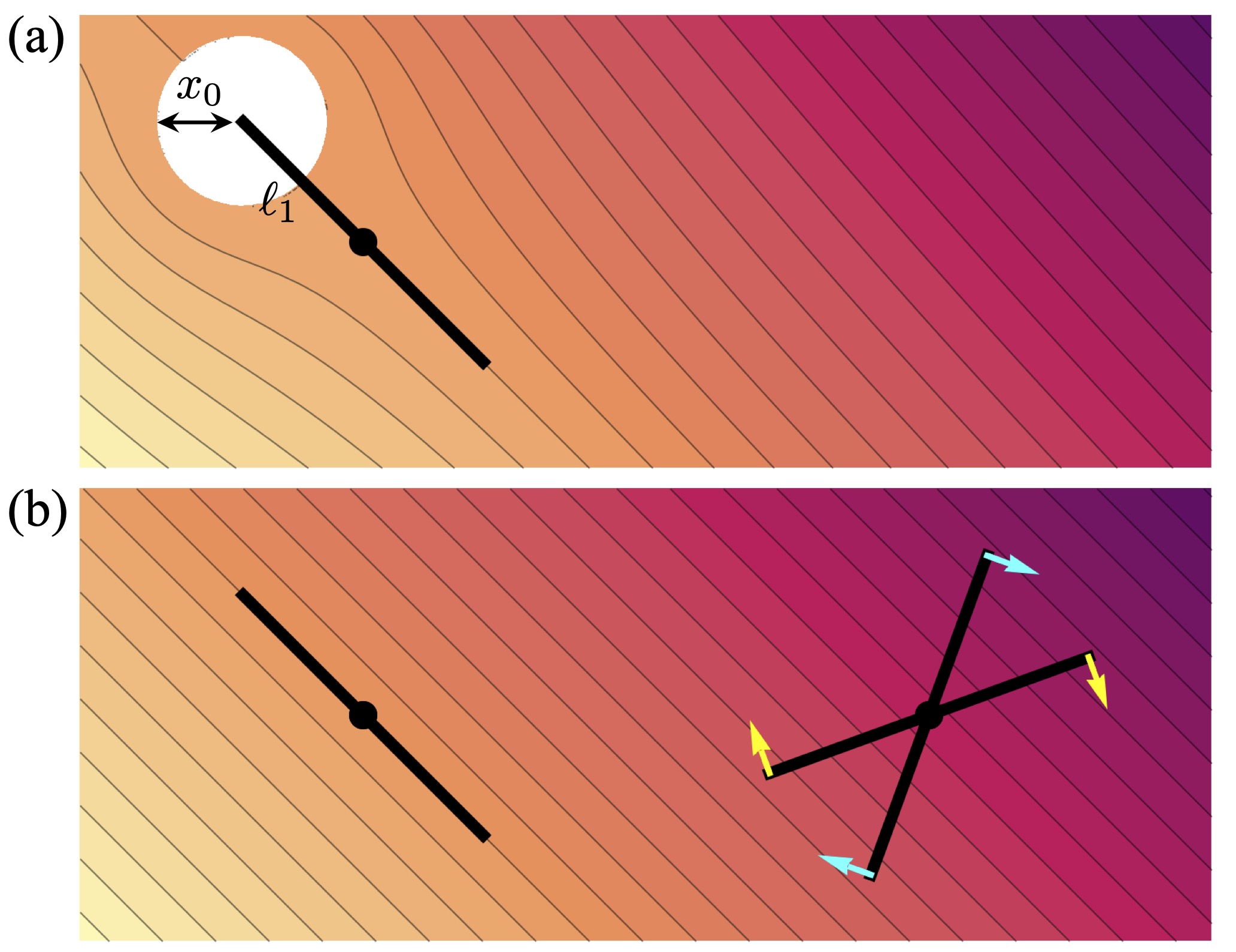}
\caption{
\textbf{Geometric origin of mixed interactions.}
(a) Contour plot of Eq.~\ref{eq:torque} for the torque $\tau_{12}$ on rotor $1$ due to a single coupling spring of rest length $x_0$. One end of the spring is mounted at $\ell_1$ on rotor $1$ and the other is held at an arbitrary location. 
(b) We may approximate $x_0 \simeq 0$ for springs mounted to a second rotor a distance $w \gg x_0$ away. 
Cyan arrows: Typical case where increasing $\theta_2$ increases the torque on rotor $1$ for uncrossed springs and decreases the torque on rotor $2$ for crossed springs. 
Yellow arrows: When the angle between the rotors exceeds $90^\circ$, this relationship inverts. 
Mixed interactions are built by inverting this relationship for some of the states. 
}
\label{fig:4}
\end{figure}

We can understand the origin of this mixed interaction by constructing maps of the torque $\tau_{12}$ on rotor $1$ due to a single coupling spring mounted at positions $\boldsymbol{x}_1$ and $\boldsymbol{x}_2$, with the origin at the pivot of rotor $1$. 
Figure~\ref{fig:4}a shows this map for $\boldsymbol{x}_1$ mounted at $\ell_1$ and with finite $x_0$. 
Setting $x_0 = 0$ modifies the map near $\boldsymbol{x}_1$ but does not change the behavior far away (Fig.~\ref{fig:4}b). 
This limiting case illustrates the perpendicular distance from rotor $1$ as the salient component of $\boldsymbol{x}_2$. 

Connecting this spring to a second rotor and keeping $|\theta| \leq 45^\circ$, an increase in $\theta_2$ increases $\tau_{12}$ for $\ell_2>0$ and decreases $\tau_{12}$ for $\ell_2<0$ (Fig.~\ref{fig:4}b, cyan arrows). 
When the angle between the rotors surpasses $90^\circ$, increasing $\theta_2$ now decreases $\tau_{12}$ for $\ell_2>0$ and increases $\tau_{12}$ for $\ell_2<0$ (Fig.~\ref{fig:4}b, yellow arrows). 
Including a matched coupling spring simply doubles the torque. 
Thus, in the large-angle state, uncrossed springs change their nature to antiferromagnetic-like, and crossed to ferromagnetic-like. 
Arranging some states to have angle differences less than $90^\circ$ and some more than $90^\circ$ thus allows one to build mixed interactions.

\subsection{Pseudoavalanches}
Mixed interactions can facilitate an exotic behavior called a pseudoavalanche, where a transition causes a chain reaction that reverts the instigating rotor its prior state. 
Consider a pair of hysterons that are brought to the transition $-- \rightarrow -+$. 
Suppose rotor $1$ feels a ferromagnetic-like coupling that causes the new state $-+$ to avalanche to $++$. 
Then, if rotor $2$ feels an antiferromagnetic-like coupling, $++$ could avalanche to $+-$. 
If $+-$ is stable, the full avalanch sequence is $-+ \rightarrow ++ \rightarrow +-$. 
While this pathway has been observed in simulations of the abstract hysteron model~\cite{Hecke21}, whether it can occur in a physical system, and how such a system could be built, have remained open questions. 

\begin{figure*}[p!]
\includegraphics[width=1.0\textwidth]{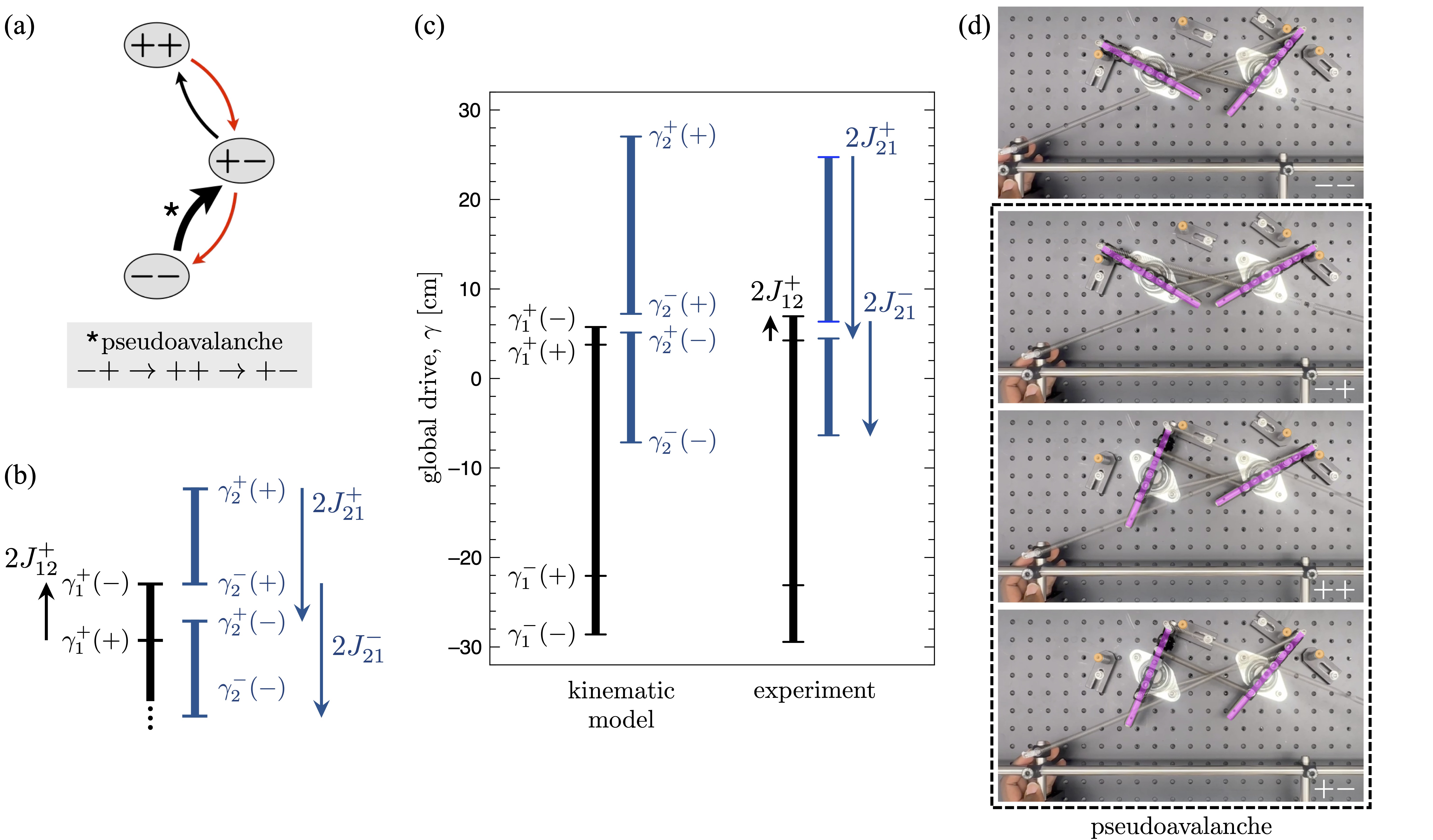}
\caption{
\textbf{Building a pseudoavalanche.}
(a) Transition graph containing a pseudoavalanche. 
This behavior requires a mixed interaction: $J_{12}^+ > 0$, $J_{21}^\pm < 0$. 
(b) One possible ordering of switching thresholds that produces this transition graph. 
The thresholds $\gamma_1^+(-)$ and $\gamma_2^-(+)$ can be in either order. 
(c) Switching thresholds consistent with this ordering, obtained from the kinematic model (Eqs.~\ref{eq:drive_exact}, \ref{eq:coupling_exact}, \ref{eq:balance}) and from the experiment. 
Mounting positions: $L=6.3$ cm, $\ell_1 = -\ell_2 = 3.2$ cm, $y=10$ cm, $w=15.24$ cm, $x_1 = 36$ cm, $x_2 = -35$ cm. 
Coupling springs: $k_{12}, k'_{12} = 0.18$, $0.19$ N/cm, $x_0, x'_0 = 4.2$, $4.7$ cm. 
Driving springs: $k_1 = k_2 = 0.037$ N/cm, $x_0 = 6.5$, $9.7$ cm. 
To reach long extensions while staying in their linear regime, each driving spring consists of two wire springs in series. 
(d) Image sequence showing the sought-after pseudoavalanche. 
Rotors and stopping posts are highlighted for clarity. 
}
\label{fig:5}
\end{figure*}

Obtaining this pseudoavalanche requires a delicate ordering of the eight switching thresholds of a hysteron pair. 
To discover this ordering we start with the transition graph in Fig.~\ref{fig:5}a that contains a pseudoavalanche, and we translate each transition into an inequality on the switching thresholds~\cite{Keim21,Hecke21}. 
First, $\gamma_2^+(-) < \gamma_1^+(-)$ enables the $-- \rightarrow -+$ transition, although this is merely to break the indexing symmetry. 
Then, $\gamma_1^+(+) < \gamma_2^+(-)$ drives the first avalanche step and $\gamma_2^+(-) < \gamma_2^-(+)$ drives the second. 
Finally, $\gamma_2^-(\pm) < \gamma_2^+(\pm)$ is a basic requirement of dissipative hysterons. 
Figure~\ref{fig:5}b shows an ordering that satisfies all the above inequalities. 

We target this behavior in our mechanical system by adjusting the rotor angles and spring mounting positions. 
We start by exploring parameter combinations on the computer, where we determine the switching thresholds numerically using Eq.~\ref{eq:balance} with exact expressions for the torques. 
One appealing feature of the rotor hysterons is that the full set of $\gamma_2^\pm$ thresholds can be moved relative to the $\gamma_1^\pm$ thresholds by adjusting $x_2 - x_1$, while preserving their internal spacing. 
We thus start by building a strong mixed interaction with a large angle difference $\theta_1^+ - \theta_1^-$, which leads to large negative $J_{21}^\pm$ that open up the crucial gap $\gamma_2^+(-) < \gamma_2^-(+)$. 
Adjusting $x_2 - x_1$ yields the ordering of Fig.~\ref{fig:5}b. 
Figure~\ref{fig:5}c shows these switching thresholds using system parameters that are readily deployed in the lab. 

We then build the system in the lab and measure all the transitions. 
The results are shown in Fig.~\ref{fig:5}c. 
Despite small differences between theory and experiment that we attribute to play in the bearing and frictional effects, the ordering in the experiment respects all the necessary inequalities. 
Supplementary Movie 1 shows that this system produces the desired pseudoavalanche. 
Importantly, once the transition to $-+$ is initiated, we hold the driving rod fixed while the two avalanche steps transpire. 
Stills from this movie are shown in Fig.~\ref{fig:5}d. 
We thus demonstrate that despite the complicated interaction they require, pseudoavalanches can be a designed feature of a simple mechanical system with few moving parts.

\section{Three-Body Interactions}\label{sec:3body}

Looking even more broadly at tunability and control of rotor hysterons, we return to Eq.~\ref{eq:model}, which models linear pairwise interactions. 
Rotor hysterons obtain such interactions when the rest length of the driving spring, $x_0$ is small. 
This linear regime is appealing for rationalizing the system behaviors, but we may look at this situation another way: What additional behaviors arise when $x_0$ is not small? 
As we now show, we may enter a parameter regime where the strength of the interaction between two hysterons is mediated by a third: $\Jij^\pm = \Jij^\pm(s_k)$.

\subsection{Tunable driving nonlinearity}

\begin{figure}[p!]
\includegraphics[width=0.4\textwidth]{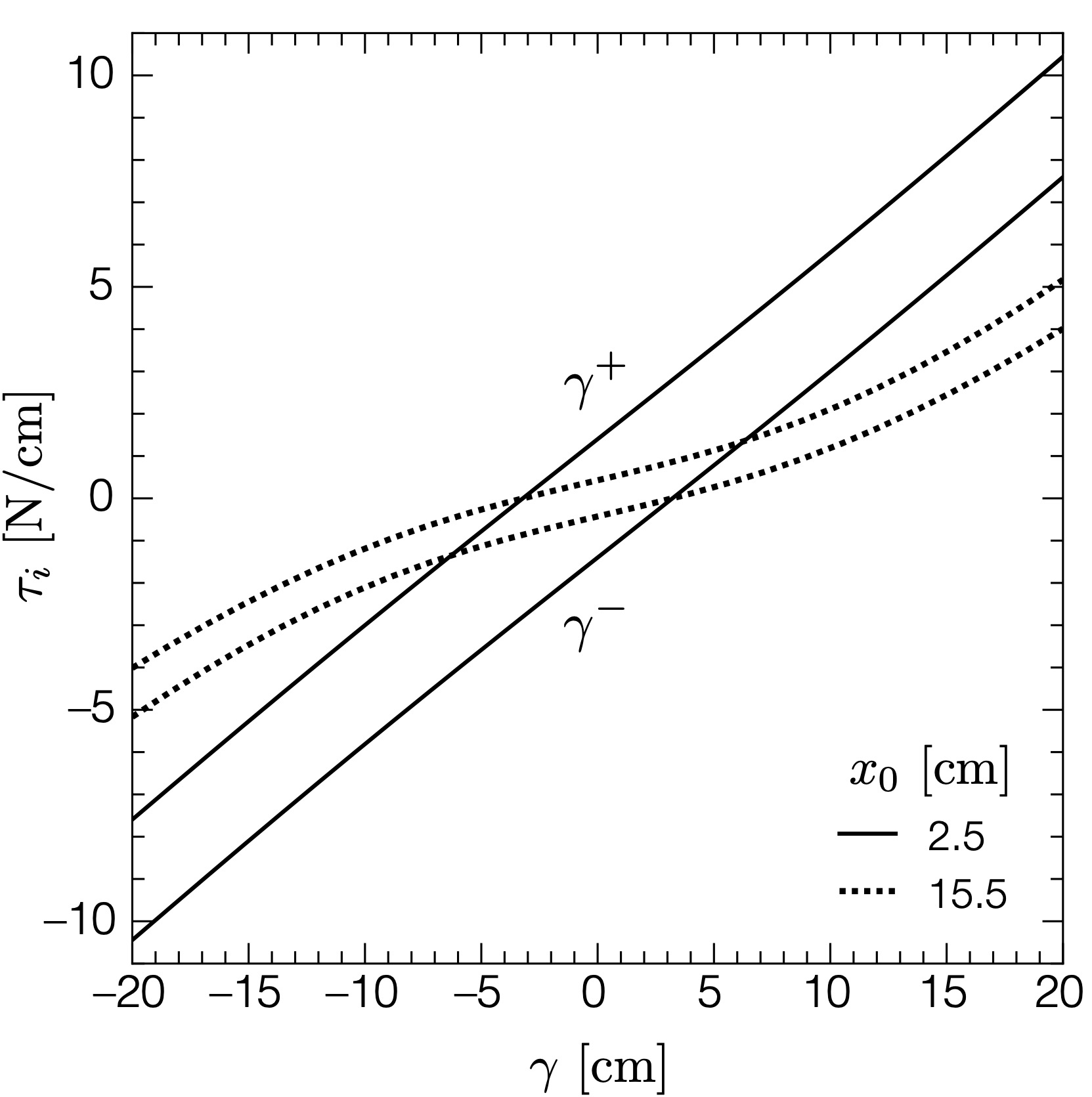}
\caption{
\textbf{Linear and nonlinear regimes of the driving spring torque.}
Equation~\ref{eq:drive_exact} for $\tau_i(\gamma)$ is plotted with two realistic values of $x_0$. 
The top curves are for $s_i = -1$ and the bottom curves are $s_i = 1$. 
When $x_0 = 2.5$ cm, the torque is approximately affine in $\gamma$. 
When $x_0 = 15.5$ cm, the torque is markedly nonlinear $\gamma$. 
This nonlinear response can lead to three-body interactions for interacting rotors. 
Other parameters values: $k_i = 0.08$ N/cm, $y=15$ cm, $L=6.31$ cm, $\theta_i^\pm = \pm 12.1^\circ$. 
}
\label{fig:6}
\end{figure}

Perhaps counterintuitively, there is a mechanism for three-body interactions that originates in the driving spring, rather than in the coupling springs that create the interaction in the first place.  
Figure~\ref{fig:6} shows the exact torque from the driving spring for a representative set of system parameters, as $x_0$ is varied over a wide range. 
When $x_0 \ll y+L$, the relationship between $\tau_i$ and $\gamma$ is approximately piecewise affine, a feature that is crucial for the linear pairwise interactions given already by Eq.~\ref{eq:Jij_from_torque}. 
In contrast, the dashed curves show the case when $x_0 = 15.5$ cm and $y+L = 21.3$ cm, leading to a significant nonlinearity between $\tau_i$ and $\gamma$. 
This nonlinearity allows a third rotor to change how one rotor interacts with a second.

\subsection{Three-rotor chain}

We consider two coupled rotors with an interaction strength $J^+_{21}$ of rotor 1 on rotor 2, as schematized in Fig.~\ref{fig:7}a. 
Recall that this interaction measures a difference of switching thresholds: $2 J^+_{21} = \gamma_2^+(s_1=-) - \gamma_2^+(s_1=+)$. 
Referring back to Eq.~\ref{eq:balance}, each of these switching thresholds stems from torque balance on rotor $2$. 
Introducing a third rotor and coupling it to rotor $2$, the two threshold torques $\tau_2(s_1 = \pm)$ will move together along the $y$-axis by an amount $\tau_{23}$. 
Because torques add linearly, the difference $\tau_2(s_1 = -) - \tau_2(s_1 = +)$ will not change. 
But, a nonlinear $\tau_2(\gamma)$ curve means that the corresponding difference on the $x$-axis, $\gamma_2^+(s_1=-) - \gamma_2^+(s_1=+)$, will depend on this shift from the third rotor, and the shift depends on $s_3$. 
This difference of the $\gamma_2^+$ defines $J_{21}^+$. 
Thus, $J^+_{21} = J^+_{21}(s_3)$.

\subsection{Magnitude of three-body interactions}

\begin{figure*}[p!]
\includegraphics[width=1.0\textwidth]{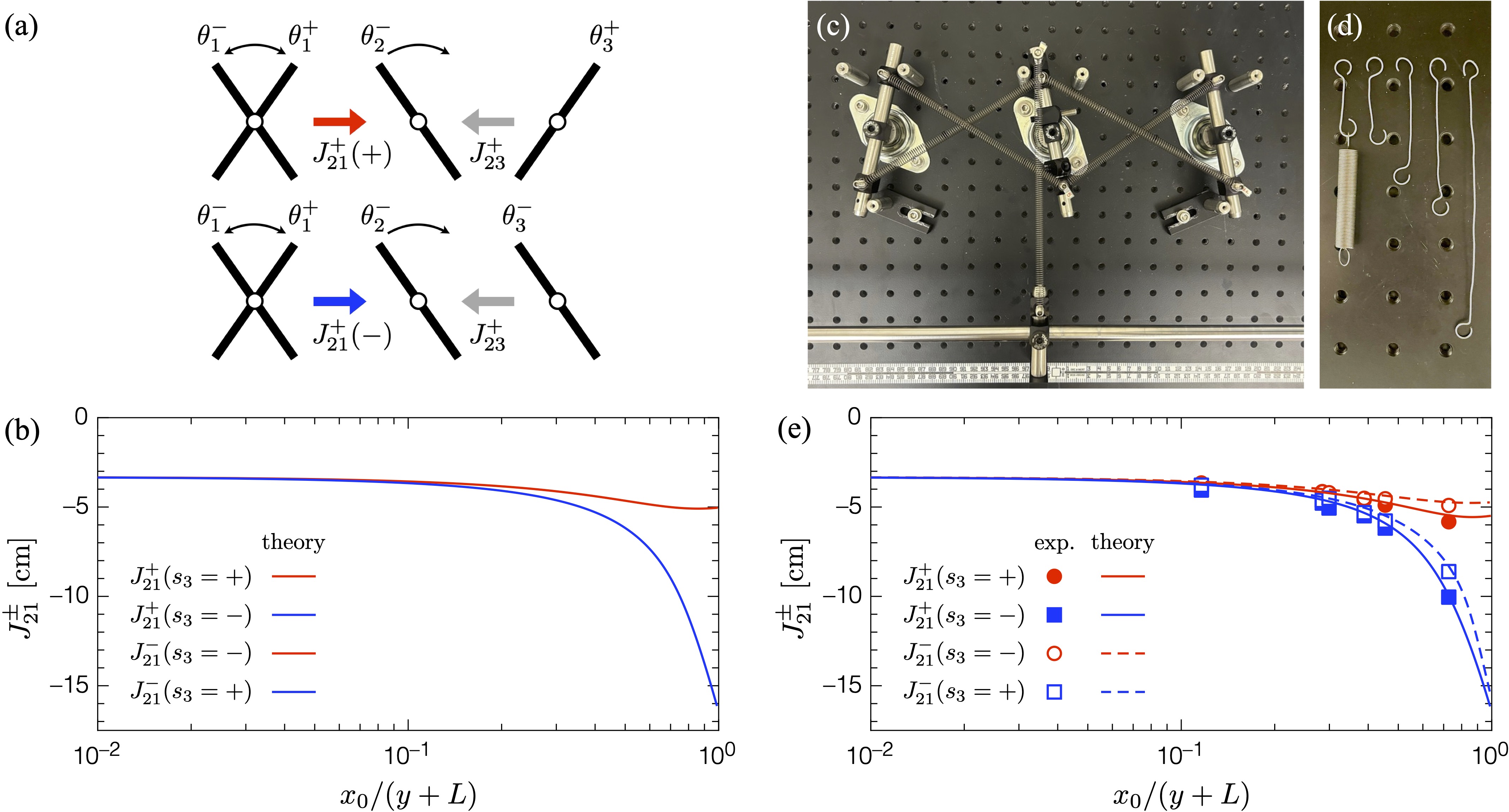}
\caption{
\textbf{Three-body interactions.}
(a) Schematic depicting $J^+_{21}$ when a third rotor is in the $(+)$ state (top) as compared with $J^+_{21}$ when a third rotor is in the $(-)$ state (bottom). 
The rotors are coupled to their neighbors in a linear chain. 
(b) Interaction strengths $J^+_{21}$ and $J^-_{21}$ for this linear chain versus the scaled driving spring rest length, $x_0/(y+L)$. 
The system shows strong three-body interactions at large $x_0$. 
Parameter values: $y=15.0$ cm, $L=6.3$ cm, $w=15.24$ cm, $\theta_i^\pm = \pm12.1^\circ$.  
Coupling springs: $k_{ij}=0.19$ N/cm, $x_0=4.1$ cm, $\ell = \pm4.73$ cm. 
Driving spring: $k_2=0.070$ N/cm. 
(c) We realize an experimental system with the same parameters. 
(d) Lengths of wire used to extend the effective rest length of the driving spring, $x_0$. 
(e) Experimental results. 
The data are captured by theory curves that use the individual measured rest length for each coupling spring, which range from $x_0=3.8$ to $4.4$ cm. 
Error bars of $\pm0.5$ cm are approximately the size of the symbols. 
}
\label{fig:7}
\end{figure*}

The next question is whether the magnitude of such three-body interactions can be appreciable. 
We begin by numerically analyzing the same three-rotor chain (Fig.~\ref{fig:7}a) coupled with crossed springs, using Eqs.~\ref{eq:drive_exact},~\ref{eq:coupling_exact}, and~\ref{eq:Jij_from_gamma}. 
This arrangement allows us to examine the interaction of rotor 1 on rotor 2 as the state of rotor 3 is varied: $J^\pm_{21}(s_3)$. 
We use realistic physical parameters so that we can later build the same setup the lab, provided in the caption to Fig.~\ref{fig:7}. 
We vary the driving spring rest length $x_0$ from $0$ up to a maximum possible value of $y + L\cos \theta_i^+$ so that the spring is never compressed. 
Figure~\ref{fig:7}b shows the two curves, $J^+_{21}(s_3=+)$ and $J^+_{21}(s_3=-)$, versus the dimensionless ratio $x_0/(y+L)$. 
The curves split at large $x_0$ by an amount comparable to the interaction strength, indicating a strong three-body interaction in that regime. 
As $x_0 \rightarrow 0$ the curves merge as the driving torque becomes piecewise affine. 

Next we examine the interactions when rotor $2$ is switching into the $(-)$ state, that is, $J^-_{21}(s_3=+)$ and $J^-_{21}(s_3=-)$. 
Interestingly, the following equality is suggested by our numerics: $J^+_{21}(s_3=\pm) = J^-_{21}(s_3=\mp)$. 
As we show later, this equality is exact and is due to a symmetry of these parameters. 
For now, we need only observe that three-body interactions are clearly present in the numerics in both $\Jij^+$ and $\Jij^-$. 

To test whether these interactions rise above the level of friction and measurement uncertainty, we build the same setup in the lab (Fig.~\ref{fig:7}c). 
We vary the rest length of the driving spring by adding lengths of stainless steel wire in series with the driving spring (Fig.~\ref{fig:7}d). 
We measure the necessary switching thresholds and then compute the interactions via Eq.~\ref{eq:Jij_from_gamma}. 
Figure~\ref{fig:7}e shows that three-body interactions are peresent. 

We notice, however, that there is a systematic shift where $J^-_{21}(+) > J^+_{21}(-)$ and $J^-_{21}(-) > J^+_{21}(+)$ consistently across the experiments. 
Revisiting the coupling spring parameters, the characteristic rest length that we used for the theory, $x_0=4.1$ cm, is the average of the four measured rest lengths, but the individual values range from $3.8$ to $4.4$ cm. 
(The measured spring constant $k_{ij}=0.19$ N/cm was consistent across the four coupling springs.) 
Figure~\ref{fig:7}e compares the experimental results to four theory curves using the individual measured spring parameters. 
We find excellent agreement, corroborating this additional higher-order effect. 

\subsection{Three-body abstract hysteron model}

Writing these three-body effects in the language of Eq.~\ref{eq:model} requires that we track the state dependence of the interaction, $\Jij=J_{ij}(S)$, which is no longer pairwise. 
Although this has advantages in that it uses familiar symbols with clear interpretations, we may pursue an alternative path by constructing an abstract hysteron model with three-body interactions tabulated in a tensor $K_{ijk}$: 
\begin{equation}\label{eq:model-3body}
\gamma_i^\pm(S) = \gamma_{i,0}^\pm - \sum_{j} \Jij^\pm s_j - \sum_{j,k} K^\pm_{ijk} s_j s_k \ ,
\end{equation}
so that $\Jij^\pm$ regains its status as a pairwise interaction. 
Comparison with Eq.~\ref{eq:model} gives a simple relationship between the two models:
\begin{equation}\label{eq:Jij-3body}
\Jij^\pm(S) = \Jij^\pm + \sum_{k} K^\pm_{ijk} s_k \ ,
\end{equation}
from which we obtain a formula for the $K^\pm_{ijk}$ in terms of the $\Jij^\pm(S)$:
\begin{equation}\label{eq:K_from_J}
\Jij^\pm(S |_{s_k=+}) - \Jij^\pm(S |_{s_k=-}) = 2 K^\pm_{ijk} \ . 
\end{equation}
This expression is analogous to Eq.~\ref{eq:Jij_from_gamma}. 
The $K^\pm_{ijk}$ are not all independent. 
For instance, $K_{ijk}^\pm = K_{ikj}^\pm$, which can be seen by converting to the switching thresholds: 
\begin{align*}
4K^\pm_{ijk} &= 2 \Jij^\pm(S |_{s_k=+}) - 2 \Jij^\pm(S |_{s_k=-}) \\
	&= \gamma_i^\pm(S |_{s_j=-, s_k=+}) - \gamma_i^\pm(S |_{s_j=+, s_k=+}) - \gamma_i^\pm(S |_{s_j=-, s_k=-}) + \gamma_i^\pm(S |_{s_j=+, s_k=-}) \\
	&= 2 J_{ik}^\pm(S |_{s_j=+}) - 2 J_{ik}^\pm(S |_{s_j=-}) \\
	&= 4K_{ikj}^\pm \ .
\end{align*}
This result and Eqs.~\ref{eq:model-3body}-\ref{eq:K_from_J} are valid for any system of interacting hysterons. 
Specializing to rotor hysterons, Eq.~\ref{eq:K_from_J} tells us that the difference between the curves in Fig.~\ref{fig:7}b is $2 K_{213}^\pm$ and the ``bare'' interaction $J_{21}^\pm$ is the average of the curves.

\section{Interaction symmetries}

Symmetry has played a central roll in the above results. 
Section~\ref{sec:Jplusminus} showed how to split $\Jij^\pm$ by breaking a symmetry. 
In Section~\ref{sec:3body}, we noted an unanticipated equality, $J^+_{21}(s_3=\pm) \simeq J^-_{21}(s_3=\mp)$ (Fig.~\ref{fig:7}b); we likewise split those values by breaking a symmetry (Fig.~\ref{fig:7}c). 
In this section we derive a general result for systems of rotor hysterons (Theorem 1), and we show how it underlies both of these behaviors. 

\subsection{Inversion symmetry of $\Jij^\pm$}

\textbf{Definition 1:} \textit{A pair of coupling springs between rotors $i$ and $j$ is said to be ``matched'' if the two coupling springs have equal spring constants and rest lengths and they are mounted at positions $\ell_i$, $\ell_i'$ on rotor $i$ with $\ell_i = -\ell_i'$ and at positions $\ell_j$, $\ell_j'$ on rotor $j$ with $\ell_j = -\ell_j'$.} 

We remind that matched springs may be in uncrossed or crossed configurations. 
In a system of rotor hysterons, we say that all coupling springs are matched if each pair is matched; the values of $k_{ij}$ and $x_0$ may vary between different pairs. 

\textbf{Definition 2:} \textit{Let $S$ be a particular system state. We denote by $\overline{S}$ the system state where all rotors have switched states.}

Depending on the context, $S$ and $\overline{S}$ may not specify the states of particular rotors, e.g.~in the expression $\Jij(S)$, the system state $S$ does not specify $s_i$ or $s_j$. 

\textbf{Theorem 1:} \textit{Consider a system of rotor hysterons where all coupling springs are matched. If $\theta_i^+ = -\theta_i^-$ for each $i$, then $\Jij^+(S) = \Jij^-(\overline{S})$.} 

\begin{figure*}[p!]
\includegraphics[width=0.85\textwidth]{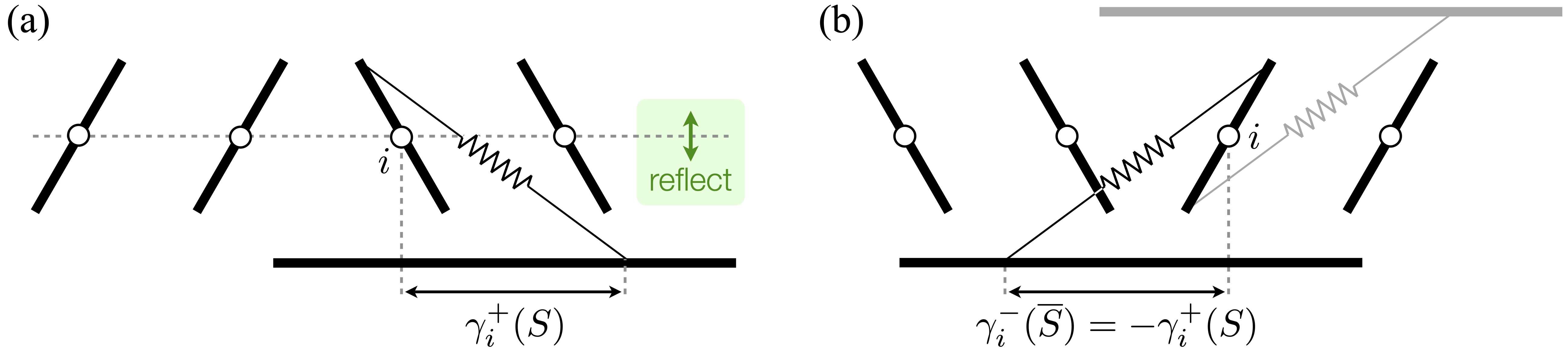}
\caption{
\textbf{Visual aid to Theorem 1.}
(a) A system in a state $S$, frozen at $\gamma_i^+(S)$. 
We let $x_i=0$, so $\gamma_i^+(S)$ is measured from the pivot of rotor $i$. 
A reflection maps $S$ to $\overline{S}$ while also preserving torque balance. 
(b) Rotating the reflected driving rod and spring $180^\circ$ about rotor $i$ sets the driving to $-\gamma_i^+(S)$ while maintaining the same torque on rotor $i$. 
This configuration corresponds to $\gamma_i^-(\overline{S})$, demonstrating their equality. 
}
\label{fig:8}
\end{figure*}

\textit{Proof:} 
Consider a system with $x_i=0$ in the state $S$, with $\gamma$ just on the verge of $\gamma_i^+(S)$. 
Reflect this system about the pivots of all the rotors (Fig.~\ref{fig:8}a). 
Because the springs are matched and the stopping posts are symmetrical, this operation flips the state of each rotor, putting the system in the state $\overline{S}$. 
The reflection preserves torque balance so that the system is now at the verge of the threshold $\gamma_i^-$, but with the driving rod on the opposite side of the rotors (Fig.~\ref{fig:8}b). 
Rotating only the driving rod and driving spring $180^\circ$ about the pivot of rotor $i$, the driving spring applies an identical torque to rotor $i$ and the driving rod recovers its usual location. 
This shows that $\gamma_i^-(\overline{S}) = -\gamma_i^+(S)$ and likewise, $\gamma_i^+(\overline{S}) = -\gamma_i^-(S)$. 
Including the effect of a nonzero $x_i$ revises this result to: $\gamma_i^\pm(\overline{S}) = -\gamma_i^\mp(S) - 2x_i$. 
By Eq.~\ref{eq:Jij_from_gamma}, 
\begin{align*}
2\Jij^+(S) &= \gamma_i^+(S |_{s_j=-}) - \gamma_i^+(S |_{s_j=+}) \\
              &= \gamma_i^-(\overline{S} |_{s_j=-}) -\gamma_i^-(\overline{S} |_{s_j=+}) \\
              &= 2\Jij^-(\overline{S})
\end{align*}

\subsection{Application to three-body interactions}

Theorem 1 immediately explains the equality $J_{21}^+(s_3=\pm) = J_{21}^-(s_3=\mp)$ observed in Fig.~\ref{fig:7}b, as well as the splitting of these curves in Fig.~\ref{fig:7}c. 
We can also recast Theorem 1 in terms of the three-body interactions $K^\pm_{ijk}$ as follows:

\textbf{Corollary 1:} \textit{Consider a system of rotor hysterons where all coupling springs are matched. If $\theta_i^+ = -\theta_i^-$ for each $i$, then $\Jij^+ = \Jij^-$ and $K_{ijk}^+ = -K_{ijk}^-$.} 

\textit{Proof:} 
By Eq.~\ref{eq:Jij-3body}, we have $J^+(S) = \Jij^+ + \sum_k K_{ijk}^+ s_k$ and $J^-(\overline{S}) = \Jij^- - \sum_k K_{ijk}^- s_k$, since $\overline{S}$ swaps all states, $s_k \rightarrow -s_k$. 
Applying Theorem 1, the difference of the two ``bare'' $\Jij$ is:
\begin{equation}
\Jij^- - \Jij^+ = \sum_k (K_{ijk}^+ + K_{ijk}^- ) s_k \ .
\end{equation}
Since this equality holds for any system state $S = \{s_k\}$, it must be the case that $K_{ijk}^+ + K_{ijk}^- = 0$ for each $k$, and hence $\Jij^- = \Jij^+$. 

This result provides a second way to explain the equality of $J_{21}^+(s_3=\pm)$ and $J_{21}^-(s_3=\mp)$ in Fig.~\ref{fig:7}b. 
Equation~\ref{eq:Jij-3body} expresses these interactions as $J_{21}^+ \pm K_{213}^+$ and $J_{21}^- \mp K_{213}^-$, respectively. 
Corollary 1 then shows that these two quantities are equal. 

\subsection{Application to pairwise interactions}

We now return to the scenario where three-body interactions are absent. 
Here the interactions are pairwise: $\Jij^\pm(S) = \Jij^\pm$, as in the general model of Eq.~\ref{eq:model}. 
Theorem 1 then gives the following corollary regarding the splitting of $\Jij^\pm$. 

\textbf{Corollary 2:} \textit{Consider a system of rotor hysterons where all coupling springs are matched and interactions are pairwise. If $\theta_i^+ = -\theta_i^-$ for each $i$, then $\Jij^+ = \Jij^- = \Jij$.} 

\textit{Proof:} Apply Theorem 1 and drop the state dependence of $\Jij^+(S)$ and $\Jij^-(\overline{S})$, since interactions are pairwise. 

This result was previously proven in Fig.~S2 of Ref.~\cite{Paulsen26} using geometric arguments; here we have shown it as a corollary to a more general result. 
We can view Section~\ref{sec:Jplusminus} through the lens of Corollary 2. 
There, we split $\Jij^\pm$ by breaking the premise $\theta_i^+ = -\theta_i^-$ (Fig.~\ref{fig:2}) and then by breaking the premise of matched springs (Fig.~\ref{fig:3}).

\section{Discussion}

We have investigated the capacity of a rotor-based mechanical hysteron platform to materialize interactions of increasing complexity. 
Our results show how to build split, mixed, and three-body interactions, and we have rationalized our results by appealing to torque balance, symmetries, and approximations. 
Perhaps most remarkable is the pseudoavalanche that was orchestrated via a mixed interaction. 
Notably, this broader set of interactions was accessed using a simple design, without any new components or degrees of freedom beyond that of Ref.~\cite{Paulsen26}. 
This approach was crucial for maintaining simple relationships between the physical parameters and the hysteron thresholds and interactions, which allowed us to explore the vast parameter space rationally, through a combination of physical and mathematical reasoning. 

We also uncovered a number of nuances within these expanded interactions. 
For instance, we found that the magnitude of the splitting of $\Jij^\pm$ can be appreciable when a neighboring $j$ has asymmetric angles, but relatively muted when the angle asymmetry is in rotor $i$ itself. 
This trend was explained by analyzing a mathematical model in the limit of small spring rest lengths. 

We considered only pair and three-body interactions, but extending our analysis to higher multibody terms should be possible. 
Indeed, if a rotor with a nonlinear $\tau_i(\gamma)$ curve is coupled to more than two hysterons, such multibody effects would arise. 
To avoid needless complexity, multibody effects can easily be limited to a subset of the rotors; only those rotors whose driving springs have large $x_0$ experience these effects. 
We also note that the highest multibody term needed is not the number of rotors, but rather the maximum number of rotors interacting with any single rotor with a nonlinear $\tau_i(\gamma)$ curve. 

Although the abstract hysteron model of Eq.~\ref{eq:model} has proven useful in capturing responses of driven disordered matter, recent work has identified behaviors not easily explained by it~\cite{Lindeman25b,Shohat25}. 
This has prompted related models~\cite{Shohat26} and alternative theoretical frameworks~\cite{Hagh26}. 
There are also examples of systems of buckled beams that possess memory and history dependence but do not fit neatly into hysteron models~\cite{Kwakernaak23}. 
Our approach in this work asks a different set of questions: Using Eq.~\ref{eq:model} as a starting point, how can this model be materialized in a mechanical structure? 
What is the broadest set of thresholds and interactions that can be accessed with a simple, replicable design and what are the physical limitations on doing so? 
Here we have demonstrated a wide range of designer interactions for mechanical hysterons built from simple components. 
Importantly, in this system the deviations from Eq.~\ref{eq:model} such as three-body effects are elective and under the designers control. 

Our results could aid the design of a wider class of machines to perform simple tasks, such as the hysteron latch that can discriminate pulse size~\cite{Lindeman25a,Paulsen26} or the accumulator that counts driving cycles~\cite{Kwakernaak23,Paulsen26}. 
One could also pursue a physical intelligence of a grander sort. 
A hysteron machine endowed with the ability to adjust its own parameters could adapt, evolve, or learn~\cite{Stern23,Altman24,Guo26} based on its environment. 
What this study offers to that pursuit is a robust, tunable platform on which to build such machines, or others still waiting to be conceived.

% Bibliography
%\bibliography{references}

\bigskip

\textbf{Acknowledgements.} 
We thank Prabal Adhikari, Martin van Hecke, Nathan Keim, and Dor Shohat for stimulating discussions. 
Funding support from NSF-DMR-2601820 is gratefully acknowledged.

%%%%%%%%%%%%
%%% FIGURES %%%
%%%%%%%%%%%%

\newpage

\end{document}